OPEN

# Constructions of Unextendible Maximally Entangled Bases in $\mathbb{C}^d \otimes \mathbb{C}^{d'}$

Gui-Jun Zhang[1], Yuan-Hong Tao[1], Yi-Fan Han[1], Xin-Lei Yong[1] & Shao-Ming Fei[2,3]

**We study unextendible maximally entangled bases (UMEBs) in $\mathbb{C}^d \otimes \mathbb{C}^{d'}$ ($d < d'$). An operational method to construct UMEBs containing $d(d' - 1)$ maximally entangled vectors is established, and two UMEBs in $\mathbb{C}^5 \otimes \mathbb{C}^6$ and $\mathbb{C}^5 \otimes \mathbb{C}^{12}$ are given as examples. Furthermore, a systematic way of constructing UMEBs containing $d(d' - r)$ maximally entangled vectors in $\mathbb{C}^d \otimes \mathbb{C}^{d'}$ is presented for $r = 1, 2, \ldots, d - 1$. Correspondingly, two UMEBs in $\mathbb{C}^3 \otimes \mathbb{C}^{10}$ are obtained.**

Quantum entanglement lies in the heart of the quantum information processing. It plays important roles in many fields such as quantum teleportation, quantum coding, quantum key distribution protocol, quantum non-locality[1–4]. Quantum teleportation, which can be used for distributed quantum learning[5] and even in organisms[6], is a essential element in quantum information processing. Maximally entangled states attract much attention due to their importance in ensuring the highest fidelity and efficiency in quantum teleportation[7]. A pure state $|\psi\rangle$ is said to be a $d \otimes d'$ ($d < d'$) maximally entangled state if and only if for an arbitrary given orthonormal basis $\{|i_A\rangle\}$ of subsystem A, there exists an orthonormal basis $\{|i_B\rangle\}$ of subsystem B such that $|\psi\rangle$ can be written as $|\psi\rangle = \frac{1}{\sqrt{d}}\sum_{i=0}^{d-1}|i_A\rangle \otimes |i_B\rangle$[8].

Nonlocality is a very useful concept in quantum mechanics[9–13] and plays an important role in Van der Waals interaction in transformation optics[14]. It is tightly related to entanglement. While, it is proven that the unextendible product bases (UPBs) reveal some nolocality without entanglement[15,16]. The UPB is a set of incomplete orthogonal product states in bipartite quantum system $\mathbb{C}^d \otimes \mathbb{C}^{d'}$ consisting of fewer than $dd'$ vectors which have no additional product states are orthogonal to each element of the set[17].

A UPB in $\mathbb{C}^3 \otimes \mathbb{C}^3$ with 5 pure states is as follows[17]:

$$|\phi_0\rangle = \frac{1}{\sqrt{2}}|0\rangle \otimes (|0\rangle - |1\rangle), \quad |\phi_1\rangle = \frac{1}{\sqrt{2}}(|0\rangle - |1\rangle) \otimes |2\rangle,$$

$$|\phi_2\rangle = \frac{1}{\sqrt{2}}|2\rangle \otimes (|1\rangle - |2\rangle), \quad |\phi_3\rangle = \frac{1}{\sqrt{2}}(|1\rangle - |2\rangle) \otimes |0\rangle,$$

$$|\phi_4\rangle = \frac{1}{3}(|0\rangle + |1\rangle + |2\rangle) \otimes (|0\rangle + |1\rangle + |2\rangle).$$

Obviously, they are all product states.

There exist a nonzero pure state $|\psi\rangle$, which is orthogonal to $|\phi_i\rangle (i = 0, 1, 2, 3, 4)$. If $|\psi\rangle$ is a product state, it can be expressed as $|\psi\rangle = (a|0\rangle + b|1\rangle + c|2\rangle) \otimes (a'|0\rangle + b'|1\rangle + c'|2\rangle)$, where $a^2 + b^2 + c^2 = a'^2 + b'^2 + c'^2 = 1$. No loss of generalization, we assume that $a, a' \neq 0$. From $|\psi\rangle$ is orthogonal to $|\phi_0\rangle$, we have $b' = a' \neq 0$. Due to $|\psi\rangle$ is orthogonal to $|\phi_4\rangle$, we can conclude that $c' \neq 0$. Because of that $|\psi\rangle$ is orthogonal to $|\phi_1\rangle$, we have $b = a \neq 0$. Owing to that $|\psi\rangle$ is orthogonal to $|\phi_3\rangle$, we obtain that $c = b \neq 0$. And $|\psi\rangle$ is orthogonal to $|\phi_2\rangle$, we get that $c' = b' \neq 0$. That is to say, $a = b = c = \frac{1}{\sqrt{3}}$ and $a' = b' = c' = \frac{1}{\sqrt{3}}$. Now $|\psi\rangle$ is equal to $|\phi_4\rangle$, instead of being orthogonal to

[1]Department of Mathematics College of Sciences, Yanbian University, Yanji, 133002, China. [2]School of Mathematics Sciences, Capital Normal University, Beijing, 100048, China. [3]Max-Planck-Institute for Mathematics in the Science, Leipzig, 04103, Germany. Correspondence and requests for materials should be addressed to Y.-H.T. (email: taoyuanhong12@126.com)

$|\phi_4\rangle$. Therefore, $|\psi\rangle$ can not be a product state. That's why the set $\{|\phi_i\rangle\}_{i=0}^4$ is described by 'unextendible' and $\{|\phi_i\rangle\}_{i=0}^4$ is a UPB in $\mathbb{C}^3 \otimes \mathbb{C}^3$.

Bravyi and Smolin[18] generalized the notion of UPB to unextendible maximally entangled bases (UMEB): a set of incomplete orthogonal maximally entangled states in bipartite quantum system $\mathbb{C}^d \otimes \mathbb{C}^{d'}$ consisting of fewer than $dd'$ vectors which have no additional maximally entangled vectors orthogonal to all of them. They state that UMEBs can be used to construct examples of states for which 1-copy entanglement of assistance (EoA) is strictly smaller than the asymptotic EoA and find quantum channels that are unital but not convex mixtures of unitary operations[18].

Let $\{|0\rangle, |1\rangle\}$ be a orthogonal base of $\mathbb{C}^2$. And $\{|0'\rangle, |1'\rangle, |2'\rangle\}$ be a orthogonal base of $\mathbb{C}^3$. Then, we present a UMEB in $\mathbb{C}^2 \otimes \mathbb{C}^3$[19]:

$$|\phi_{1,2}\rangle = \frac{1}{\sqrt{2}}(|0\rangle \otimes |0'\rangle \pm |1\rangle \otimes |1'\rangle),$$

$$|\phi_{3,4}\rangle = \frac{1}{\sqrt{2}}(|0\rangle \otimes |1'\rangle \pm |1\rangle \otimes |0'\rangle).$$

Obviously, they are all maximally entangled states.

If a nonzero pure state $|\psi\rangle$ is orthogonal to $|\phi_i\rangle (i = 1, 2, 3, 4)$, it's sure that $|\psi\rangle = (a|0\rangle + b|1\rangle) \otimes |2'\rangle$, where $a^2 + b^2 = 1$. In other words, $|\psi\rangle$ must be a product state, rather than maximally entangled states. Hence $\{|\phi_i\rangle\}_{i=0}^4$ is a UMEB in $\mathbb{C}^2 \otimes \mathbb{C}^3$.

The number of the vectors in a UMEB is less than the dimension of the bipartite system space. Therefore a UMEB in $\mathbb{C}^d \otimes \mathbb{C}^{d'}$ containing $n$ maximally entangled vectors is usually expressed as a $n$-number UMBE, when $n$ is smaller than $dd'$. Chen and Fei[19] provided a way to construct $d^2$-member UMEBs in $\mathbb{C}^d \otimes \mathbb{C}^{d'} \left(\frac{d'}{2} < d < d'\right)$. Later, Nan *et al.*[20] and Li *et al.*[21] constructed two sets of UMEBs in $\mathbb{C}^d \otimes \mathbb{C}^{d'}$ $(d < d')$ independently. Wang *et al.*[22] put forward a method of constructing UMEBs in $\mathbb{C}^{qd} \otimes \mathbb{C}^{qd}$ from that in $\mathbb{C}^d \otimes \mathbb{C}^d$, and gave a 30-member UMEB in $\mathbb{C}^6 \otimes \mathbb{C}^6$. They proved that there exist UMEBs in $\mathbb{C}^d \otimes \mathbb{C}^d$ except for $d = p$ or $2p$, where $p$ is a prime and $p = 3$ mod 4. They also presented a 23-member UMEB in $\mathbb{C}^5 \otimes \mathbb{C}^5$ and a 45-member UMEB in $\mathbb{C}^7 \otimes \mathbb{C}^7$[23]. Then Guo[24–26] proposed a scenario of constructing UMEBs via the space decomposition, which improves the previous work about UMEBs.

In this paper, we give two methods of constructing UMEBs in $\mathbb{C}^d \otimes \mathbb{C}^{d'} (d \leq d')$. In Sec. 2 we first recall some basic notions and lemmas about UMEB and space decomposition. In Sec. 3 we give an operational method to construct $d(d' - 1)$-number UMEB and then present explicit constructions of UMEBs in $\mathbb{C}^5 \otimes \mathbb{C}^6$ and $\mathbb{C}^5 \otimes \mathbb{C}^{12}$. In Sec. 4 we present an approach of systematically constructing $d(d' - r)$-member UMEBs in $\mathbb{C}^d \otimes \mathbb{C}d'$ for $r = 1, 2, \ldots, d - 1$, and give two examples in $\mathbb{C}^3 \otimes \mathbb{C}^{10}$. We summarize in Sec. 5.

## Preliminaries

Throughout this paper, we assume that $d < d'$. Let us first recall some basic notions and lemmas[18,19,24]. Let $\{|k\rangle\}$ and $\{|\ell'\rangle\}$ be the standard computational bases of $\mathbb{C}^d$ and $\mathbb{C}^{d'}$, respectively, and $\{|\phi_i\rangle\}_{i=1}^{dd'}$ an orthonormal basis of $\mathbb{C}^d \otimes \mathbb{C}^{d'}$. Let $M_{d\times d'}$ be the Hilbert space of all $d \times d'$ complex matrices equipped with the inner product defined by $\langle A|B\rangle = Tr(A^\dagger B)$ for any $A, B \in M_{d\times d'}$. If $\{A_i\}_{i=1}^{dd'}$ constitutes a Hilbert-Schmidt basis of $M_{d\times d'}$, where $\langle A_i|A_j\rangle = d\delta_{ij}$, then there is a one-to-one correspondence between $\{|\phi_i\rangle\}$ and $\{A_i\}$ as follows[25,26]:

$$|\phi_i\rangle = \sum_{k,\ell} a_{kl}^{(i)}|k\rangle|\ell'\rangle \in \mathbb{C}^d \otimes \mathbb{C}^{d'} \quad \Leftrightarrow \quad A_i = [\sqrt{d}\, a_{k\ell}^{(i)}] \in M_{d\times d'},$$

$$Sr(|\phi_i\rangle) = rank(A_i), \qquad \langle \phi_i|\phi_j\rangle = \frac{1}{d}Tr(A_i^\dagger A_j), \tag{1}$$

where $Sr(|\phi_i\rangle)$ denotes the Schmidt number of $|\phi_i\rangle$. Obviously, $|\phi_i\rangle$ is a maximally entangled pure state in $C^d \otimes C^{d'}$ iff $(d)^{1/2}A_i$ is a $d \times d'$ singular-value-1 matrix (a matrix whose singular values all equal to 1).

A basis $\{|\phi_i\rangle\}_{i=1}^{dd'}$ constituted by maximally entangled states in $C^d \otimes C^{d'}$ is called a maximally entangled basis (MEB) of $C^d \otimes C^{d'}$. A set of pure states $\{|\phi_i\rangle\}_{i=1}^{n} \in C^d \otimes C^{d'}$ with the following conditions is called an unextendible maximally entangled basis (UMEB)[18,19]:

(i) $|\phi_i\rangle$, $i = 1, 2, 3 \ldots n$ are all maximally entangled states.
(ii) $\langle \phi_i|\phi_j\rangle = \delta_{ij}, \ i, j = 1, 2, 3, \ldots, n.$
(iii) $n < dd'$, and if a pure state $|\psi\rangle$ satisfies that $\langle \phi_i|\psi\rangle = 0$, $i = 1, 2, 3\ldots n$, then $|\psi\rangle$ can not be maximally entangled.

A Hilbert-Schmidt basis $\{A_i\}_{i=1}^{dd'}$ constituted by single-value-1 matrices in $M_{d\times d'}$ is called single-value-1 Hilbert-Schmidt basis (SV1B) of $M_{d\times d'}$. A set of $d \times d'$ matrices $\{A_i\}_{i=1}^{n}$ with the following conditions is called unextendible singular-value-1 Hilbert-Schmidt basis (USV1B) of $M_{d\times d'}$[24]:

(i) $A_i$, $i = 1, 2, 3 \ldots n$ are all single-value-1 matrices.
(ii) $Tr(A_i^\dagger A_j) = d\delta_{ij}$, $i, j = 1, 2, 3\ldots n$.
(iii) $n < dd'$, and if a matrix $X$ satisfies that $Tr(X^\dagger A_i) = 0$, $i = 1, 2, 3\ldots n$, then $X$ can not be a single-value-1 matrix.

It is obvious that $\{A_i\}_{i=1}^{dd'}$ is an SV1B of $M_{d\times d'}$ iff $\{|\phi_i\rangle\}_{i=1}^{dd'}$ is a MEB of $C^d\otimes C^{d'}$, and $\{A_i\}_{i=1}^{n}$ is a USV1B of $M_{d\times d'}$ iff $\{|\phi_i\rangle\}_{i=1}^{n}$ is a UMEB of $C^d\otimes C^{d'}$. Therefore, for convenience, we may just call an SV1B $\{A_i\}_{i=1}^{dd'}$ of $M_{d\times d'}$ an MEB $\{|\phi_i\rangle\}_{i=1}^{dd'}$ of $\mathbb{C}^d\otimes\mathbb{C}^{d'}$, and call a USV1B $\{A_i\}_{i=1}^{n}$ of $M_{d\times d'}$ a UMEB $\{|\phi_i\rangle\}_{i=1}^{n}$ of $\mathbb{C}^d\otimes\mathbb{C}^{d'}$.

In deriving our main results, we need the following lemma in ref.[24].

**Lemma 1**. Let $M_{d\times d'} = M_1\oplus M_1^{\perp}$. If $\{|\phi_i\rangle\}$ is a MEB in $M_1$ and $\{|\psi_i\rangle\}$ is a UMEB in $M_1^{\perp}$, then $\{|\phi_i\rangle\}\cup\{|\psi_i\rangle\}$ is a UMEB in $M$[24]. If $\{|\phi_i\rangle\}$ is a MEB in $M_1$ and $M_1^{\perp}$ contains no single-value-1 matrix (maximally entangled state), then $\{|\phi_i\rangle\}$ is a UMEB in $M$.

## $d(d'-1)$-member UMEBs in $\mathbb{C}^d\otimes\mathbb{C}^{d'}$

In this section, we will establish a flexible method to construct $d(d'-1)$-member UMEBs in $\mathbb{C}^d\otimes\mathbb{C}^{d'}$.

**Theorem 1**. Let $M_{d\times d'}$ be the Hilbert space of all $d\times d'$ complex matrices. If $V$ is a subspace of $M_{d\times d'}$ such that each matrix in $V$ is a $d\times d'$ matrix ignoring $d$ entries which occupy different rows and $N$ columns with $N < d$, then there exists a $d(d'-1)$-member MEB in $V$, as well as a $d(d'-1)$-member UMEB in $M_{d\times d'}$.

**Proof**. Without loss of generality, we can always assume the ignored $d$ entries in $V$ only occupy the former $N$ columns. Let $b_i$, $i = 0, 1, \ldots, d-1$, denote the column number of the ignored element in the $i$-th row. Obviously, $b_{i+1} - b_i = 0$ or 1.

Denote

$$C(k,\ l) = \begin{cases} 1, & l = b_k;\\ 0, & otherwise.\end{cases} \tag{2}$$

We can construct $d(d'-1)$ pure states in $\mathbb{C}^d\otimes\mathbb{C}^{d'}$ as follows,

$$|\phi'_{j,n}\rangle = \frac{1}{\sqrt{d}}\sum_{m=0}^{d-1}\omega_d^{nm}|m\rangle|t_{mj}\rangle,\ j = 0, 1, \ldots, d'-2; n = 0, 1, \ldots, d-1, \tag{3}$$

where $\omega_d = e^{\frac{2\pi\sqrt{-1}}{d}}$, and

$$t_{mj} = \begin{cases} j+1, & m = 0;\\ t_{m-1,j} + 1\oplus_{d'} C(m,\ t_{m-1,j}+1), & m = 1, 2, \ldots, d-1,\end{cases} \tag{4}$$

$p\oplus_{d'}m$ denotes $(p+m)$ mod $d'$.

Next, we prove that all the states in (3) constitute an MEB in $V$.

(i) Maximally entangled.

If $C(m, t_{m-1,j}\oplus_{d'}1) = 0$ for any $m$, it is obvious that $t_{mj}\neq t_{m'j}$ for $m\neq m'$.

If $C(m, t_{m-1,j}\oplus_{d'}1) = 1$ for some $m\neq 0$, from the definition of $t_{mj}$ one has $t_{m-1,j}\neq b_{m-1}$. Note that $b_m - b_{m-1} = 0$ or 1, then $t_{m-1,j} = b_{m-1}\oplus_{d'}1$.

From the definition of $t_{mj}$, we also have $C(k+1, t_{kj}+1) = 0$ for $k\neq m-1$. Hence

$$t_{kj} = \begin{cases} t_{m-1,j}\ominus_{d'}(m-1-k), & 0\leqslant k < m-1;\\ t_{mj}\oplus_{d'}(k-m), & k\geqslant m.\end{cases} \tag{5}$$

where $p\ominus_{d'}m$ denotes $(p-m)$ mod $d'$. In particular,

$$t_{0j} = t_{m-1,j}\ominus_{d'}(m-1),\qquad t_{d-1,j} = t_{mj}\oplus_{d'}(d-1-m). \tag{6}$$

Then

$$\begin{aligned} t_{d-1,j} - t_{0j} &= t_{mj} \oplus_{d'} (d-1-m) - t_{m-1,j} \ominus_{d'} (m-1) \\ &= d - 2 + (t_{mj} - t_{m-1,j}) \\ &= d < d'. \end{aligned} \tag{7}$$

Hence $t_{mj} \neq t_{m'j}$ for $m \neq m'$. Namely, the states $|\phi'_{j,n}\rangle$ in (3) are all maximally entangled.

(ii) Orthogonality.

We first show that $|t_{mj}\rangle = |t_{mj'}\rangle$ if and only if $j = j'$.

Obviously, $t_{mj} = t_{mj'}$ for $j = j'$. If $j \neq j'$, without loss of generality, let $j' > j$. It is easy to show that $t_{mj} \neq t_{mj'}$ when $t_{m-1,j} \neq t_{m-1,j'}$. Otherwise, from the definition of $t_{mj}$ we have $t_{mj} = t_{m-1,j'} \ominus_{d'} C(m, t_{m-1,j} \oplus 1)$ when $C(m, t_{m-1,j} \oplus_{d'} 1) = 1$. Note that $t_{m-1,j'} = b_{m-1} \ominus_{d'} 1$ when $C(m, t_{m-1,j} \oplus_{d'} 1) = 1$, as proved in (i). Therefore, $t_{m-1,j'} = b_{m-1}$, which contradicts to the definition of $t_{mj}$. Furthermore, $t_{mj} \neq t_{mj'}$ when $t_{0j} \neq t_{0j'}$. Therefore,

$$\begin{aligned} \langle \phi'_{j,n} | \phi'_{j',n'} \rangle &= \frac{1}{d} \sum_{m=0}^{d-1} \overline{\omega^{n'm}} \omega^{nm} \langle t_{mj} | t_{mj'} \rangle \\ &= \frac{1}{d} \sum_{m=0}^{d-1} \omega^{(n-n')m} \delta_{jj'} \\ &= \delta_{nn'} \delta_{jj'}. \end{aligned} \tag{8}$$

Thus, the $d(d'-1)$ states $\{|\phi_{j,n}\rangle\}$ in (3) constitutes an MEB in $V$. Furthermore, there exist no MEBs in $V^{\perp}$ because $N < d$. Hence $\{|\phi_{j,n}\rangle\}$ is a UMEB in $M_{d\times d'}$, as well as in $\mathbb{C}^d \otimes \mathbb{C}^{d'}$.

**Example 1**. Constructing two UMEBs in $C^5 \otimes C^6$, where as $V = \begin{pmatrix} 0 & 0 & 0 & * & 0 & 0 \\ 0 & * & 0 & 0 & 0 & 0 \\ 0 & 0 & 0 & * & 0 & 0 \\ 0 & 0 & 0 & 0 & 0 & * \\ 0 & 0 & 0 & * & 0 & 0 \end{pmatrix}$.

We can get the following matrix $V'$ by using suitable unitary transformation on $V$,

$$V' = PVQ = \begin{pmatrix} * & 0 & 0 & 0 & 0 & 0 \\ * & 0 & 0 & 0 & 0 & 0 \\ * & 0 & 0 & 0 & 0 & 0 \\ 0 & * & 0 & 0 & 0 & 0 \\ 0 & 0 & * & 0 & 0 & 0 \end{pmatrix},$$

where

$$P = \begin{pmatrix} 1 & 0 & 0 & 0 & 0 \\ 0 & 0 & 1 & 0 & 0 \\ 0 & 0 & 0 & 0 & 1 \\ 0 & 1 & 0 & 0 & 0 \\ 0 & 0 & 0 & 1 & 0 \end{pmatrix}, \quad Q = \begin{pmatrix} 0 & 0 & 0 & 1 & 0 & 0 \\ 0 & 1 & 0 & 0 & 0 & 0 \\ 0 & 0 & 0 & 0 & 1 & 0 \\ 1 & 0 & 0 & 0 & 0 & 0 \\ 0 & 0 & 0 & 0 & 0 & 1 \\ 0 & 0 & 1 & 0 & 0 & 0 \end{pmatrix}.$$

According to Theorem 1, we first construct an MEB $\{|\phi'_j\rangle\}_{j=1}^{25}$ in $V'$, i.e. a UMEB in $C^5 \otimes C^6$ as follows:

$$\begin{cases} |\phi'_{1,2,3,4,5}\rangle &= \frac{1}{\sqrt{5}}(|01'\rangle + \alpha|12'\rangle + \alpha^2|23'\rangle + \alpha^3|34'\rangle + \alpha^4|45'\rangle), \\ |\phi'_{6,7,8,9,10}\rangle &= \frac{1}{\sqrt{5}}(|02'\rangle + \alpha|13'\rangle + \alpha^2|24'\rangle + \alpha^3|35'\rangle + \alpha^4|40'\rangle), \\ |\phi'_{11,12,13,14,15}\rangle &= \frac{1}{\sqrt{5}}(|03'\rangle + \alpha|14'\rangle + \alpha^2|25'\rangle + \alpha^3|30'\rangle + \alpha^4|41'\rangle), \\ |\phi'_{16,17,18,19,20}\rangle &= \frac{1}{\sqrt{5}}(|04'\rangle + \alpha|15'\rangle + \alpha^2|21'\rangle + \alpha^3|32'\rangle + \alpha^4|43'\rangle), \\ |\phi'_{21,22,23,24,25}\rangle &= \frac{1}{\sqrt{5}}(|05'\rangle + \alpha|11'\rangle + \alpha^2|22'\rangle + \alpha^3|33'\rangle + \alpha^4|44'\rangle), \end{cases} \tag{9}$$

where $\alpha = 1, \omega_5, \omega_5^2, \omega_5^3, \omega_5^4$.

By inverse unitary transformation $|\phi_j\rangle = (P^{-1} \otimes Q^{-1})|\phi_j'\rangle$, we get the following MEB $\{|\phi_j\rangle\}_{j=1}^{25}$ in $V$, i.e., another UMEB in $C^5 \otimes C^6$:

$$\begin{cases} |\phi_{1,2,3,4,5}\rangle = \frac{1}{\sqrt{5}}(|01'\rangle + \alpha|25'\rangle + \alpha^2|40'\rangle + \alpha^3|12'\rangle + \alpha^4|34'\rangle), \\ |\phi_{6,7,8,9,10}\rangle = \frac{1}{\sqrt{5}}(|05'\rangle + \alpha|20'\rangle + \alpha^2|42'\rangle + \alpha^3|14'\rangle + \alpha^4|33'\rangle), \\ |\phi_{11,12,13,14,15}\rangle = \frac{1}{\sqrt{5}}(|00'\rangle + \alpha|22'\rangle + \alpha^2|44'\rangle + \alpha^3|13'\rangle + \alpha^4|31'\rangle), \\ |\phi_{16,17,18,19,20}\rangle = \frac{1}{\sqrt{5}}(|02'\rangle + \alpha|24'\rangle + \alpha^2|41'\rangle + \alpha^3|15'\rangle + \alpha^4|30'\rangle), \\ |\phi_{21,22,23,24,25}\rangle = \frac{1}{\sqrt{5}}(|04'\rangle + \alpha|21'\rangle + \alpha^2|45'\rangle + \alpha^3|10'\rangle + \alpha^4|32'\rangle), \end{cases} \quad (10)$$

where $\alpha = 1, \omega_5, \ \omega_5^2, \ \omega_5^3, \ \omega_5^4$.

**Remark 1**. Actually both (9) and (10) are UMEBs in $\mathbb{C}^5 \otimes \mathbb{C}^6$. However, they are different although they can be unitarily transformed to each other. We will reveal the difference in the following example.

**Example 2**. Constructing a UMEB in $\mathbb{C}^5 \otimes \mathbb{C}^{12}$, whereas

$$V = (V_1|V_2) = \begin{pmatrix} 0 & 0 & 0 & * & 0 & 0 & * & 0 & 0 & 0 & 0 & 0 \\ 0 & 0 & 0 & * & 0 & 0 & 0 & 0 & 0 & * & 0 & 0 \\ 0 & * & 0 & 0 & 0 & 0 & * & 0 & 0 & 0 & 0 & 0 \\ 0 & 0 & 0 & * & 0 & 0 & 0 & 0 & 0 & * & 0 & 0 \\ 0 & 0 & 0 & * & 0 & 0 & * & 0 & 0 & 0 & 0 & 0 \end{pmatrix}. \quad (11)$$

One can easily get the following simple formations $V'_1$ and $V'_2$ from $V_1$ and $V_2$ by elementary transformation respectively:

$$V_1' = P_1 V_1 Q_1 = \begin{pmatrix} * & 0 & 0 & 0 & 0 & 0 \\ 0 & * & 0 & 0 & 0 & 0 \\ 0 & * & 0 & 0 & 0 & 0 \\ 0 & * & 0 & 0 & 0 & 0 \\ 0 & * & 0 & 0 & 0 & 0 \end{pmatrix}, \quad V'_2 = P_2 V_2 Q_2 = \begin{pmatrix} * & 0 & 0 & 0 & 0 & 0 \\ * & 0 & 0 & 0 & 0 & 0 \\ * & 0 & 0 & 0 & 0 & 0 \\ 0 & * & 0 & 0 & 0 & 0 \\ 0 & * & 0 & 0 & 0 & 0 \end{pmatrix}. \quad (12)$$

Then following Theorem 1 we can construct the following UMEBs $\{|\phi_{j'}\rangle\}_{j=1}^{25}$ and $\{|\psi_{j'}\rangle\}_{j=1}^{25}$ in $V'_1$ and $V'_2$ respectively:

$$\begin{cases} |\phi'_{1,2,3,4,5}\rangle = \frac{1}{\sqrt{5}}(|01'\rangle + \alpha|12'\rangle + \alpha^2|23'\rangle + \alpha^3|34'\rangle + \alpha^4|45'\rangle), \\ |\phi'_{6,7,8,9,10}{}'\rangle = \frac{1}{\sqrt{5}}(|02'\rangle + \alpha|13'\rangle + \alpha^2|24'\rangle + \alpha^3|35'\rangle + \alpha^4|40'\rangle), \\ |\phi'_{11,12,13,14,15}{}'\rangle = \frac{1}{\sqrt{5}}(|03'\rangle + \alpha|14'\rangle + \alpha^2|25'\rangle + \alpha^3|30'\rangle + \alpha^4|42'\rangle), \\ |\phi'_{16,17,18,19,20}{}'\rangle = \frac{1}{\sqrt{5}}(|04'\rangle + \alpha|15'\rangle + \alpha^2|20'\rangle + \alpha^3|32'\rangle + \alpha^4|43'\rangle), \\ |\phi'_{21,22,23,24,25}{}'\rangle = \frac{1}{\sqrt{5}}(|05'\rangle + \alpha|10'\rangle + \alpha^2|22'\rangle + \alpha^3|33'\rangle + \alpha^4|44'\rangle), \end{cases} \quad (13)$$

$$\begin{cases} |\psi'_{1,2,3,4,5}\rangle = \frac{1}{\sqrt{5}}(|07'\rangle + \alpha|18'\rangle + \alpha^2|29'\rangle + \alpha^3|3, 10'\rangle + \alpha^4|4, 11'\rangle), \\ |\psi'_{6,7,8,9,10}\rangle = \frac{1}{\sqrt{5}}(|08'\rangle + \alpha|19'\rangle + \alpha^2|2, 10'\rangle + \alpha^3|3, 11'\rangle + \alpha^4|46'\rangle), \\ |\psi'_{11,12,13,14,15}\rangle = \frac{1}{\sqrt{5}}(|09'\rangle + \alpha|1, 10'\rangle + \alpha^2|2, 11'\rangle + \alpha^3|3, 6'\rangle + \alpha^4|48'\rangle), \\ |\psi'_{16,17,18,19,20}\rangle = \frac{1}{\sqrt{5}}(|0, 10'\rangle + \alpha|1, 11'\rangle + \alpha^2|27'\rangle + \alpha^3|38'\rangle + \alpha^4|49'\rangle), \\ |\psi'_{21,22,23,24,25}\rangle = \frac{1}{\sqrt{5}}(|0, 11'\rangle + \alpha|17'\rangle + \alpha^2|28'\rangle + \alpha^3|39'\rangle + \alpha^4|4, 10'\rangle) \end{cases} \quad (14)$$

where $\alpha = 1, \omega_5, \ \omega_5^2, \ \omega_5^3, \ \omega_5^4$.

By inverse transformation $|\phi_j\rangle = (P_1^{-1} \otimes Q_1^{-1})|\phi'_j\rangle$ and $|\psi_j\rangle = (P_2^{-1} \otimes Q_2^{-1})|\psi'_j\rangle$, we can obtain the following UMEBs $\{|\phi_j\rangle\}_{j=1}^{25}$ and $\{|\psi_j\rangle\}_{j=1}^{25}$ in $V_1$ and $V_2$, respectively,

$$\begin{cases} |\phi_{1,2,3,4,5}\rangle = \frac{1}{\sqrt{5}}(|01'\rangle + \alpha|12'\rangle + \alpha^2|23'\rangle + \alpha^3|34'\rangle + \alpha^4|45'\rangle), \\ |\phi_{6,7,8,9,10}\rangle = \frac{1}{\sqrt{5}}(|02'\rangle + \alpha|13'\rangle + \alpha^2|24'\rangle + \alpha^3|35'\rangle + \alpha^4|40'\rangle), \\ |\phi_{11,12,13,14,15}\rangle = \frac{1}{\sqrt{5}}(|03'\rangle + \alpha|14'\rangle + \alpha^2|25'\rangle + \alpha^3|30'\rangle + \alpha^4|42'\rangle), \\ |\phi_{16,17,18,19,20}\rangle = \frac{1}{\sqrt{5}}(|04'\rangle + \alpha|15'\rangle + \alpha^2|20'\rangle + \alpha^3|32'\rangle + \alpha^4|43'\rangle), \\ |\phi_{21,22,23,24,25}\rangle = \frac{1}{\sqrt{5}}(|05'\rangle + \alpha|10'\rangle + \alpha^2|22'\rangle + \alpha^3|33'\rangle + \alpha^4|44'\rangle), \end{cases} \quad (15)$$

$$\begin{cases} |\psi_{1,2,3,4,5}\rangle & = \frac{1}{\sqrt{5}}(|07'\rangle + \alpha|18'\rangle + \alpha^2|29'\rangle + \alpha^3|3,\ 10'\rangle + \alpha^4|4,\ 11'\rangle), \\ |\psi_{6,7,8,9,10}\rangle & = \frac{1}{\sqrt{5}}(|08'\rangle + \alpha|19'\rangle + \alpha^2|2,\ 10'\rangle + \alpha^3|3,\ 11'\rangle + \alpha^4|46'\rangle), \\ |\psi_{11,12,13,14,15}\rangle & = \frac{1}{\sqrt{5}}(|09'\rangle + \alpha|1,\ 10'\rangle + \alpha^2|2,\ 11'\rangle + \alpha^3|36'\rangle + \alpha^4|48'\rangle), \\ |\psi_{16,17,18,19,20}\rangle & = \frac{1}{\sqrt{5}}(|0,\ 10'\rangle + \alpha|1,\ 11'\rangle + \alpha^2|27'\rangle + \alpha^3|38'\rangle + \alpha^4|49'\rangle), \\ |\psi_{21,22,23,24,25}\rangle & = \frac{1}{\sqrt{5}}(|0,\ 11'\rangle + \alpha|17'\rangle + \alpha^2|28'\rangle + \alpha^3|39'\rangle + \alpha^4|4,\ 10'\rangle). \end{cases} \tag{16}$$

Thus, $\{|\phi_j\rangle\}\cup\{|\psi_j\rangle\}$ constitutes a UMEB in $\mathbb{C}^5 \otimes \mathbb{C}^{12}$ with $V$ in (8). However, neither $(P_1^{-1} \otimes Q_1^{-1})$ nor $(P_2^{-1} \otimes Q_2^{-1})$ can transform $\left\{\left|\psi'_j\right\rangle\right\} \cup \left\{\left|\phi'_j\right\rangle\right\}$ to $\left\{\left|\phi_j\right\rangle\right\} \cup \left\{\left|\psi_j\right\rangle\right\}$, which shows the difference between (9) and (10).

## *d*(*d*′ − *r*)-member UMEBs in $\mathbb{C}^d \otimes \mathbb{C}^{d'}$

In this section, we construct UMEBs consisting of fewer elements in $\mathbb{C}^d \otimes \mathbb{C}^{d'}$. The following theorem provides a systematic way of constructing $d(d' - r)$-member UMEBs in $\mathbb{C}^d \otimes \mathbb{C}^{d'}$, $r = 1, 2, \ldots, d - 1$, that is to say, it presents $d - 1$ constructions of UMEB in $\mathbb{C}^d \otimes \mathbb{C}^{d'}$.

**Theorem 2**. Let $d' = \sum_{i=1}^{s} a_i + r$, where $s \geqslant 1,\ a_i \geqslant d,\ \ 0 < r < d$. Then the following vectors constitute a $d(d' - r)$-member UMEB in $C^d \otimes C^{d'}$:

$$|\phi_{l,j,n}\rangle = \frac{1}{\sqrt{d}}\sum_{m=0}^{d-1}\omega_d^{nm}|m\rangle|b_j + (l_{j+i} \oplus_{a_{j+1}} m)\rangle, \quad {}_{j+1} = 0,\ \ 1,\ \ \cdots,\ a_{j+1} - 1, \tag{17}$$

where $b_j = \sum_{k=1}^{j} a_k$; $j = 0, 1,\ldots, s - 1$; $n = 0, 1 \ldots, d - 1$.

**Proof.** (i) It is obvious that $|\phi_{l,j,n}\rangle$ in (16) are all maximally entangled.

(ii) Orthogonality,

$$\begin{aligned} \langle\phi_{l,j,n}|\phi_{l',j',n'}\rangle &= \frac{1}{d}\sum_{m=0}^{d-1}\overline{\omega_d^{n'm}}\omega_d^{nm}\langle b_j + (l_{j+1} \oplus_{a_{j+1}} m)|b_{j'} + (l'_{j'+1} \oplus_{a_{j'+1}} m)\rangle \\ &= \frac{1}{d}\sum_{m=0}^{d-1}\omega_d^{(n-n')m}\delta_{jj'}\langle l_{j+1} \oplus_{a_{j+1}} m|l'_{j'+1} \oplus_{a_{j'+1}} m\rangle \\ &= \frac{1}{d}\sum_{m=0}^{d-1}\omega_d^{(n-n')m}\delta_{jj'}\delta_{ll'} \\ &= \delta_{nn'}\delta_{jj'}\delta_{ll'}. \end{aligned} \tag{18}$$

(iii) Denote $M_1$ the $d \otimes (d' - n)$ matrix space, a subspace of $M_{d\times d'}$. Since the number of $\{|\phi_{l,j,n}\rangle\}$ in (17) equals to the dimension of $M_1$, $\{|\phi_{l,j,n}\rangle\}$ is an MEB of $M_1$. Moreover, since $M_1^{\perp}$ is a $d \times r$ matrix space and $r < d$, there contains no UMEB in $M_1^{\perp}$. From Lemma 1, $\{|\phi_{l,j,n}\rangle\}$ is a UMEB of $\mathbb{C}^d \otimes \mathbb{C}^{d'}$.

**Example 3**. UMEBs in $\mathbb{C}^3 \otimes \mathbb{C}^{10}$.

Obviously, $10 = 4 + 5 + 1$ or $10 = 4 + 4 + 2$. According to Theorem 2, we can construct the following 27-number UMEB (19) and 24-number UMEB (20) in $\mathbb{C}^3 \otimes \mathbb{C}^{10}$ respectively.

$$\begin{cases} |\phi_{1,2,3}\rangle & = \frac{1}{\sqrt{3}}(|00'\rangle + \alpha|11'\rangle + \alpha^2|22'\rangle), \\ |\phi_{4,5,6}\rangle & = \frac{1}{\sqrt{3}}(|01'\rangle + \alpha|12'\rangle + \alpha^2|23'\rangle), \\ |\phi_{7,8,9}\rangle & = \frac{1}{\sqrt{3}}(|02'\rangle + \alpha|13'\rangle + \alpha 2|20'\rangle), \\ |\phi_{10,11,12}\rangle & = \frac{1}{\sqrt{3}}(|03'\rangle + \alpha|10'\rangle + \alpha^2|21'\rangle), \\ |\phi_{13,14,15}\rangle & = \frac{1}{\sqrt{3}}(|04'\rangle + \alpha|15'\rangle + \alpha^2|26'\rangle), \\ |\phi_{16,17,18}\rangle & = \frac{1}{\sqrt{3}}(|05'\rangle + \alpha|16'\rangle + \alpha^2|27'\rangle), \\ |\phi_{19,20,21}\rangle & = \frac{1}{\sqrt{3}}(|06'\rangle + \alpha|17'\rangle + \alpha^2|28'\rangle), \\ |\phi_{22,23,24}\rangle & = \frac{1}{\sqrt{3}}(|07'\rangle + \alpha|18'\rangle + \alpha^2|24'\rangle), \\ |\phi_{25,26,27}\rangle & = \frac{1}{\sqrt{3}}(|08'\rangle + \alpha|14'\rangle + \alpha^2|25'\rangle), \end{cases} \tag{19}$$

and

$$\begin{cases} |\phi_{1,2,3}\rangle &= \frac{1}{\sqrt{3}}(|00'\rangle + \alpha|11'\rangle + \alpha^2|22'\rangle), \\ |\phi_{4,5,6}\rangle &= \frac{1}{\sqrt{3}}(|01'\rangle + \alpha|12'\rangle + \alpha^2|23'\rangle), \\ |\phi_{7,8,9}\rangle &= \frac{1}{\sqrt{3}}(|02'\rangle + \alpha|13'\rangle + \alpha^2|20'\rangle), \\ |\phi_{10,11,12}\rangle &= \frac{1}{\sqrt{3}}(|03'\rangle + \alpha|10'\rangle + \alpha^2|21'\rangle), \\ |\phi_{13,14,15}\rangle &= \frac{1}{\sqrt{3}}(|04'\rangle + \alpha|15'\rangle + \alpha^2|26'\rangle), \\ |\phi_{16,17,18}\rangle &= \frac{1}{\sqrt{3}}(|05'\rangle + \alpha|16'\rangle + \alpha^2|27'\rangle), \\ |\phi_{19,20,21}\rangle &= \frac{1}{\sqrt{3}}(|06'\rangle + \alpha|17'\rangle + \alpha^2|24'\rangle), \\ |\phi_{22,23,24}\rangle &= \frac{1}{\sqrt{3}}(|07'\rangle + \alpha|14'\rangle + \alpha^2|25'\rangle), \end{cases} \tag{20}$$

where $\alpha = 1, \omega_3, \ \omega_3^2$.

**Remark 2**. Theorem 2 gives a very large number of UMEBs in $C^d \otimes C^{d'}$, which is more than all the previous numbers. For example, the 27-number UMEB (19) and 24-number UMEB (20) in Example 3 are only two kinds of UMEBs in $\mathbb{C}^3 \otimes \mathbb{C}^{10}$. Actually according to Theorem 2, there are five more kinds of UMEBs in $\mathbb{C}^3 \otimes \mathbb{C}^{10}$, since 10 $= 3 + 5 + 2$, $10 = 3 + 6 + 1$, $10 = 3 + 3 + 3 + 1$, $10 = 8 + 2$ and $10 = 9 + 1$.

**Remark 3**. Theorem 2 in ref.[21] is a special case of the above Theorem 2 at $d' = a_1 + r$. Theorem 1 in ref.[20] and Theorem 1 in ref.[21] are both special cases of our Theorem 1, where all the $a_i$ are equal.

## Conclusion

We have provided new constructions of unextendible maximally entangled bases in arbitrary bipartite spaces $\mathbb{C}^d \otimes \mathbb{C}^{d'}$. We have presented a systematic way of constructing $d(d' - 1)$-member UMEB in $\mathbb{C}^d \otimes \mathbb{C}^{d'}$, and constructed two different UMEBs in $\mathbb{C}^3 \otimes \mathbb{C}^6$ and $\mathbb{C}^3 \otimes \mathbb{C}^{12}$ respectively. We have established a flexible method to construct $d(d - r)$-number UMEBs in $\mathbb{C}^d \otimes \mathbb{C}^{d'}$, $r = 1, 2, \ldots, d - 1$. Namely, we have presented more than $d - 1$ constructions of UMEBs in $\mathbb{C}^d \otimes \mathbb{C}^{d'}$. Such generalized the main results in ref.[21] and ref.[20]. We have also shown 27-number UMEB and 24-number UMEB in $\mathbb{C}^3 \otimes \mathbb{C}^{10}$, respectively.

## References

1. Horodecki, R., Horodecki, P., Horodecki, M. & Horodecki, K. Quantum entanglement. *Reviews of Modern Physics.* **81**, 865–942 (2002).
2. Barenco, A. & Ekert, A. K. Dense Coding Based on Quantum Entanglement. *Journal of Modern Optics.* **42**, 1253–1259 (1995).
3. Curty, M., Lewenstein, M. & Tkenhaus, N. L. Entanglement as a precondition for secure quantum key distribution. *Physical Review Letters.* **92**, 217903 (2009).
4. Zheng, S. B. Quantum nonlocality for a three-particle nonmaximally entangled state without inequalities. *Physical Review A.* **66**, 90–95 (2002).
5. Sheng, Y. B. & Zhou, L. Distributed secure quantum machine learning. *Science Bulletin.* **62**, 1025–1029 (2017).
6. Li, T. C. & Yin, Z. Q. Quantum superposition, entanglement, and state teleportation of a microorganism on an electromechanical oscillator. *Science Bulletin.* **61**, 163–171 (2016).
7. Adhikari, S., Majumdar, A. S., Roy, S., Ghosh, B. & Nayak, N. Teleportation via maximally and non-maximally entangled mixed states. *Quantum Information & Computation.* **10**, 398–419 (2010).
8. Peres, A. Quantum Theory: Concepts and Methods. *Kluwer Academic Publishers.* **28**, 131–135 (1995).
9. Cao, D. Y., Liu, B. H. & Wang, Z. Multiuser-to-multiuser entanglement distribution based on 1550 nm polarization-entangled photons. *Science bulletin.* **60**, 1128–1132 (2015).
10. Wang, Z., Zhang, C., Huang, Y. F., Liu, B. H. & Li, C. F. Experimental verification of genuine multipartite entanglement without shared reference frames. *Science Bulletin.* **61**, 714–719 (2016).
11. Guo, W. J., Fan, D. H. & Wei, L. F. Experimentally testing Bells theorem based on Hardys nonlocal ladder proofs. *Science China Physics, Mechanics & Astronomy.* **58**, 024201 (2015).
12. Meng, H. X., Cao, H. X., Wang, W. H., Chen, L. & Fan, Y. J. Continuity of the robustness of contextuality and the contextuality cost of empirical mode ls. *Science China Physics, Mechanics & Astronomy.* **59**(4), 640303 (2016).
13. Silveiro, I., Ortega, J. M. P. & Abajo, F. J. G. D. Quantum nonlocal effects in individual and interacting graphene nanoribbons. *Light: Science & Applications.* **4**, e241 (2015).
14. Zhao, R. K., Luo, Y. & Pendry, J. B. Transformation optics applied to van der Waals interactions. *Science bulletin.* **61**, 59–67 (2016).
15. DiVincenzo, D. P., Mor, T., Shor, P. W., Smolin, J. A. & Terhal, B. M. Unextendible product bases, uncompletable product bases and bound entanglement. *Commun. Math. Phys.* **238**, 379 (2003).
16. Rinaldis, S. D. Distinguishability of complete and unextendible product bases. *Phys. Rev. A.* **70**, 022309 (2004).
17. Bennett, C. H. *et al.* Unextendible product bases and bound entanglement. *Phys. Rev. Lett.* **82**, 5385 (1999).
18. Bravyi, S. & Smolin, J. A. Unextendible maximally entangled bases. *Phys. Rev. A.* **84**, 042306 (2011).
19. Chen, B. & Fei, S. M. Unextendible maximally entangled bases and mutually unbiased bases. *Phys. Rev. A.* **88**, 034301 (2013).
20. Nan, H., Tao, Y. H., Li, L. S. & Zhang, J. Unextendible Maximally Entangled Bases and Mutually Unbiased Bases in $C^d \otimes C^{d'}$. *International Journal of Theoretical Physics.* **54**, 927 (2015).
21. Li, M. S., Wang, Y. L., Fei, S. M. & Zheng, Z. J. Unextendible maximally entangled bases in $C^d \otimes C^{d'}$. *Phys. Rev. A.* **89**, 062313 (2014).
22. Wang, Y. L., Li, M. S. & Fei, S. M. Unextendible maximally entangled bases in $C^d \otimes C^{d'}$. *Phys. Rev. A.* **90**, 034301 (2014).
23. Wang, Y. L., Li, M. S. & Fei, S. M. Connecting the UMEB in $C^d \otimes C^{d'}$ with partial Hadamard matrices. *Quantum Information Processing.* **16**, 84 (2017).
24. Guo., Y. Constructing the unextendible maximally entangled basis from the maximally entangled basis. *Phys. Rev. A.* **94**, 052302 (2016).
25. Guo, Y., Du, S. P., Li, X. L. & Wu, S. J. Entangled bases with fixed Schmidt number. *Journal of Physics A: Mathematical and Theoretical.* **48**, 245301 (2015).
26. Guo, Y., Jia, Y. P. & Li, X. L. Multipartite unextendible entangled basis. *Quantum Inf Process.* **14**, 3553 (2015).

## Acknowledgements
The work is supported by the NSFC under number 11361065, 11675113, 11761073.

## Author Contributions
G.-J.Z. and Y.-H.T. wrote the main manuscript text. All of the authors reviewed the manuscript.

## Additional Information
**Competing Interests:** The authors declare no competing interests.